\documentclass[aps, prl, reprint, superscriptaddress, floatfix, nolongbibliography]{revtex4-2}

\usepackage{amsmath, amssymb, amsfonts, bm}
\usepackage{graphicx}
\usepackage{float}
\usepackage{color}
\usepackage[normalem]{ulem}
\usepackage[resetlabels]{multibib}

\usepackage[
colorlinks=true, linkcolor=blue, citecolor=blue, urlcolor=blue, 
setpagesize=false
]{hyperref}

\newcommand{\nuc}{N_{\mathrm{UC}}}
\newcommand{\szt}{\langle S_{\mathrm{tot}}^{z} \rangle}
\newcommand{\phic}[2]{\phi_{\mathrm{c}#2}^{#1}}

\newcommand{\ket}[1]{|{#1}\rangle}
\newcommand{\bra}[1]{\langle{#1}|}

\begin{document}

\title{Higher-Order Topological Mott Insulator on the Pyrochlore Lattice} 

\author{Yuichi Otsuka}
\email{otsukay@riken.jp}
\affiliation{Computational Materials Science Research Team, 
RIKEN Center for Computational Science (R-CCS), 
Kobe, Hyogo 650-0047, Japan}
\affiliation{Quantum Computational Science Research Team, 
RIKEN Center for Quantum Computing (RQC), 
Wako, Saitama 351-0198, Japan}

\author{Tsuneya Yoshida}
\affiliation{Graduate School of Pure and Applied Sciences, 
University of Tsukuba, Tsukuba, Ibaraki 305-8571, Japan}
\affiliation{Department of Physics,
University of Tsukuba, Tsukuba, Ibaraki 305-8571, Japan}

\author{Koji Kudo}
\affiliation{Department of Physics,
University of Tsukuba, Tsukuba, Ibaraki 305-8571, Japan}

\author{Seiji Yunoki}
\affiliation{Computational Materials Science Research Team,
RIKEN Center for Computational Science (R-CCS), 
Kobe, Hyogo 650-0047, Japan}
\affiliation{Quantum Computational Science Research Team, 
RIKEN Center for Quantum Computing (RQC), 
Wako, Saitama 351-0198, Japan}
\affiliation{Computational Condensed Matter Physics Laboratory, 
RIKEN,
Wako, Saitama 351-0198, Japan}
\affiliation{Computational Quantum Matter Research Team, 
RIKEN Center for Emergent Matter Science (CEMS), 
Wako, Saitama 351-0198, Japan}

\author{Yasuhiro Hatsugai}
\affiliation{Graduate School of Pure and Applied Sciences, 
University of Tsukuba, Tsukuba, Ibaraki 305-8571, Japan}
\affiliation{Department of Physics,
University of Tsukuba, Tsukuba, Ibaraki 305-8571, Japan}

\date{\today}

\begin{abstract}
We provide the first unbiased evidence for a higher-order topological 
Mott insulator in three dimensions by numerically exact quantum Monte 
Carlo simulations.
This insulating phase is adiabatically connected to a third-order 
topological insulator in the noninteracting limit, which features gapless modes 
around the corners of the pyrochlore lattice and is characterized by 
a $\mathbb{Z}_{4}$ spin-Berry phase. 
The difference between the correlated and non-correlated topological
phases is that in the former phase the gapless corner modes emerge only 
in spin excitations being Mott-like.
We also show that 
the topological phase transition from the third-order topological Mott 
insulator to the usual Mott insulator occurs when the bulk spin gap solely
closes.
\end{abstract}


\maketitle

\section{Introduction} 
Nontrivial topological properties and many-body effects are 
the two major subjects in modern condensed matter physics.
In a system involving these two subjects, 
obtaining knowledge of a wave function, often required for characterizing 
topological properties, is difficult and demanding because of the many-body 
nature.
In such a situation, the adiabatic-connection approach and the notion of 
bulk-edge correspondence can still provide smoking-gun evidence for an 
interacting topological phase.

The topological Mott insulator (TMI) is a novel state of matter
in which nontrivial topological properties and correlation effects
coexist~\cite{Hohenadler_review_2013, Rachel_review_2018, note-TMI}.
Such a state was first proposed by Pesin and Balents as one of 
possible ground states for Ir-based pyrochlore oxides~\cite{Pesin_NatPhys2009}.
Among various interesting issues originated in their proposal,
gapless surface spin-only excitations in the TMI are intriguing,
since it is in sharp contrast to the case of the usual topological 
insulators where gapless edge excitations appear in the single-particle 
spectrum~\cite{Hasan_RMP2010, Moore_Nature2010, Qi_RMP2011}.
Namely, in the TMI, 
the bulk-edge (boundary)
correspondence~\cite{Hatsugai_PRL1993, Hatsugai_PRB1993}, 
one of the most distinguished and ubiquitous properties of the 
topological insulators, is generalized by the correlation effect.
Soon after the proposal, intensive studies have examined the 
possibility of TMI in several condensed matter 
systems~\cite{Yamaji_PRB2011, Hohenadler_PRL2011, Yu_PRL2011, Zheng2011,
Yoshida_PRB2012, Tada_PRB2012, Yoshida_PRB2013, Yoshida_PRL2014, Yoshida_PRB2016, 
Bi_PRL2017, Zhang_PRB2016},
yielding concrete evidences for one~\cite{Yoshida_PRL2014}
and two~\cite{Yoshida_PRB2016, Bi_PRL2017, Zhang_PRB2016} 
dimensional cases.
However, the TMI in three dimensions (3D) has not yet been fully explored,
partly because of lack of reliable methods to study 
the correlated systems in 3D
such as the complicated model considered for the 
Ir oxides~\cite{Pesin_NatPhys2009}.

On the one hand, recently, 
another type of unconventional topological insulators, 
a higher-order topological insulator (HOTI),
has been attracting increasing 
interest~\cite{Benalcazar_Science2017, Schindler_SciAdv2018}.
The $n$th-order topological insulator in $d$-dimensions features 
gapless excitations around its ($d-n$)-dimensional boundaries.
Thus, also in HOTI, the bulk-edge correspondence is generalized.
The studies of the HOTI have not always been 
material-oriented~\cite{Schindler_NatPhys2018, Yue_NatPhys2019,
Gray_NanoLett2019, 
Liu_NanoLett2019, Sheng_PRL2019, Chen_PRL2020},
but also have covered a wide range of 
models~\cite{Benalcazar_Science2017, Schindler_SciAdv2018, 
Hashimoto_PRB2017, Benalcazar_PRB2017, Song_PRL2017, 
Fukui_PRB2018, Ezawa_PRL2018a, Ezawa_PRB2018a, 
Langbehn_PRL2017, Ezawa_PRL2018b, Ezawa_PRB2018b, 
Khalaf_PRB2018, Liu_PRL2019, Calugaru2019, 
Araki_PRB2019, Araki_PRR2020, Mizoguchi_JPSJ2019, 
You_PRB2018, Kudo_PRL2019, 
Dubinkin_PRB2019, Bibo_PRB2020, 
Peng_PRB2019, Guo_arXiv2020} 
and 
experimental 
setups~\cite{Imhof_NatPhys2018, Peterson_Nature2018, 
Serra-Garcia_Nature2018, 
Ni_NatMater2019, Xue_NatMater2019, Xue_PRL2019, 
Kempkes_NatMater2019, Weiner_SciAdv2020}.
Among them, three of the present authors proposed
a tailored model to investigate the correlation effects on the HOTI in $d=2$
and found a correlated topological state dubbed as
a higher-order topological Mott insulator (HOTMI),
in which gapless corner modes emerge only in spin excitations~\cite{Kudo_PRL2019}.

In this study, 
we present unbiased numerical evidence for a HOTMI in 3D
by constructing a repulsive Hubbard model with spin-dependent hoppings 
on the pyrochlore lattice.
Our results 
support the TMI in 3D 
in the sense that, both in the TMI and the HOTMI,
the nontrivial bulk topological property manifests itself 
in the edge states only through the spin channel.
As in the case of the kagome lattice~\cite{Kudo_PRL2019},
the repulsive Hubbard model with spin-dependent hoppings 
on the pyrochlore lattice
can be mapped into the attractive Hubbard model by the particle-hole 
transformation, and hence we can utilize a quantum Monte Carlo 
(QMC) method for the correlated model in 3D without facing the negative-sign 
problem. 
We show that the on-site interaction ($U$) added to the HOTI closes neither
the charge nor spin gap in the bulk, which suggests that the higher-order topology 
characterized by a $\mathbb{Z}_{4}$ spin-Berry phase~\cite{Hatsugai_EPL2011}
in the HOTI is adiabatically preserved in the $U>0$ phase.
As for the properties around the boundaries,
the characteristic gapless corner modes are found only in the spin sector.
These results indicate that the $U>0$ phase, next to the HOTI, is 
the third-order topological Mott insulator in 3D.
We also show that the gapless corner modes 
disappear 
when the bulk spin gap
vanishes at a phase boundary between the HOTMI and the Mott insulator (MI).

\section{Results}

\subsection{Model}
\begin{figure}[tb]
 \centering
 \includegraphics[width=0.49\linewidth]{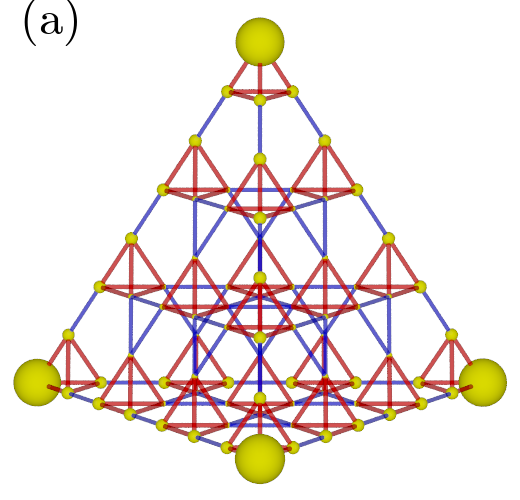}
 \includegraphics[width=0.49\linewidth]{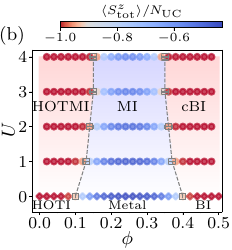}
 \caption{\label{fig:main}%
 (a) Pyrochlore lattice for $L=4$ with the open boundary conditions.
 The small upward tetrahedron represents the unit cell.
 Transfer integral for the intra (inter) unit cell 
 indicated by red (blue)
 is $t_{\bigtriangleup}$ ($t_{\bigtriangledown}$).
 Enhancement of the local moment by the on-site interaction $U=1$, i.e.,
 $\langle m_{i}^{2} \rangle_{U=1} - \langle m_{i}^{2} \rangle_{U=0}$,
 is shown by the radius of the yellow spheres
 for $\phi=0.08$ and $T=0.08$.
 (b) Ground-state phase diagram as function of $\phi$ and $U$.
 The color of the two symbols, 
 circles for $U>0$ and diamonds at $U=0$, 
 represents the value of $\langle S_{\mathrm{tot}}^{z} \rangle /N_{\mathrm{UC}}$.
 }
\end{figure}

We study the spinful interacting model on the pyrochlore lattice.
The Hamiltonian is described by
\begin{equation}
\mathcal{H} = 
  \mathcal{H}_{t}^{\bigtriangleup} + \mathcal{H}_{t}^{\bigtriangledown}
  +  \mathcal{H}_{U}
  - \mu \, N
  - h   \, S_{\mathrm{tot}}^{z}, 
  \label{eq:Hamiltonian}
\end{equation}
with
\begin{equation}
 \mathcal{H}_{t}^{\Gamma} = 
  - t_{\Gamma} 
  \sum_{i,j \in \Gamma}
  \sum_{\alpha, \beta = \uparrow, \downarrow}
  \left(
   c_{i \alpha}^{\dagger}  \sigma_{\alpha \beta}^{z}  c_{j \beta}
   + \mathrm{h.c.}
  \right)
  \label{eq:kinetic_term}
\end{equation}
and
\begin{equation}
 \mathcal{H}_{U} = 
  U \sum_{i} 
  \left( n_{i \uparrow  } - \frac{1}{2} \right)
  \left( n_{i \downarrow} - \frac{1}{2} \right),
  \label{eq:Hubbard_term}
\end{equation}
where 
$c_{i \alpha}^{\dagger}$ creates an electron 
with spin $\alpha$ $(=\uparrow, \downarrow)$ at site $i$, 
$\sigma_{\alpha \beta}^{z}$ is the $z$-component of Pauli matrix, and 
$n_{i \alpha}=c_{i \alpha}^{\dagger}c_{i \alpha}$ is a number operator.
Since we consider the model in the grand canonical ensemble,
we explicitly include the terms for
a chemical potential $\mu$ and a magnetic field $h$,
which are coupled to the total number of the electrons,
$N=\sum_{i \alpha} n_{i \alpha}$,
and the total magnetization,
$S_{\mathrm{tot}}^{z}=\sum_{i}\left(n_{i \uparrow} - n_{i \downarrow}\right)/2$.
In the kinetic part $\mathcal{H}_{t}^{\Gamma}$
with $\Gamma=\bigtriangleup$ and $\bigtriangledown$,
$t_{\bigtriangleup}$ and $t_{\bigtriangledown}$
denotes the transfer integrals for the intra and inter unit cell,
respectively [see Fig.~\ref{fig:main}(a)].
Their relative ratio is parameterized by $\phi$
with $0\le\phi\le1/2$
as  $t_{\bigtriangleup}   = t \sin\left( \phi \, \pi \right)$ 
and $t_{\bigtriangledown} = t \cos\left( \phi \, \pi \right)$.
Here, $t$ is chosen as an energy unit, namely, $t=1$.
The Hubbard term of Eq.~(\ref{eq:Hubbard_term}) represents
the repulsive ($>0$) on-site interaction.

The system preserves a certain type of particle-hole symmetry defined by
a transformation of
$c_{i \uparrow  }  \rightarrow \tilde{c}_{i \downarrow}^{\dagger}$
and
$c_{i \downarrow}  \rightarrow \tilde{c}_{i \uparrow  }^{\dagger}$,
under which the Hamiltonian is invariant for $\mu=0$.
The number of electrons with spin up, $\langle n_{i \uparrow} \rangle$, 
where $\langle \cdots \rangle$ denotes an expectation value defined below,
is related to that for spin down in the transformed Hamiltonian as
$\langle n_{i \uparrow} \rangle = 1 - \langle \tilde{n}_{i \downarrow} \rangle$.
Since the invariant Hamiltonian trivially yields the same expectation value,
$\langle \tilde{n}_{i \downarrow} \rangle = \langle n_{i \downarrow} \rangle$,
the system is half filled, i.e., 
$\langle n_{i \uparrow} \rangle + \langle n_{i \downarrow} \rangle =1$, 
at $\mu=0$, which is the case we consider in this study.

The 
unusual ingredient in our model 
would be $\sigma_{\alpha \beta}^{z}$ in Eq.~(\ref{eq:kinetic_term}),
which simply leads to the spin-dependent transfer integrals.
Such a modification of the model was proposed
in the previous study~\cite{Kudo_PRL2019} for the kagome lattice
to induce the bulk gap in the single-particle spectrum,
necessary for realizing the topological phases.
In this study, $\sigma_{\alpha \beta}^{z}$ is also crucial 
for allowing the sign-problem-free QMC calculations.

We employ the finite-temperature auxiliary-field quantum Monte Carlo 
method~\cite{Blankenbecler_PRD1981,Hirsch_PRB1985,White_PRB1989,Scalettar_PRB1991,Assaad_Book2008}.
An expectation value of a physical operator $\mathcal{O}$ at a finite temperature $T$
is calculated in the grand canonical ensemble as
$\langle \mathcal{O} \rangle = \frac{1}{Z}
\mathrm{Tr} \left( \mathcal{O} \ e^{-\beta \mathcal{H}} \right)$,
where $Z=\mathrm{Tr}\left( e^{-\beta \mathcal{H}}\right)$ is the partition function,
and $\beta=1/T$ denotes an inverse temperature.
To be convinced that our model is sign-problem free, 
let us consider a partial particle-hole transformation,
$c_{i \uparrow  }  \rightarrow  \tilde{c}_{i \uparrow  }$ and
$c_{i \downarrow}  \rightarrow  \tilde{c}_{i \downarrow}^{\dagger}$,
which maps the Hamiltonian into the following form 
(excluding a constant term):
\begin{align}
\mathcal{H}  \rightarrow \tilde{\mathcal{H}} 
=&
- \sum_{\Gamma = \bigtriangleup, \bigtriangledown}
  \sum_{i,j \in \Gamma} 
  \sum_{\alpha = \uparrow, \downarrow} t_{\Gamma}
\left(
\tilde{c}_{i \alpha}^{\dagger} \tilde{c}_{j \alpha}
+
\mathrm{h.c.}
\right)  
\nonumber \\
& - U \sum_{i} 
\left( \tilde{n}_{i \uparrow  } - \frac{1}{2} \right)
\left( \tilde{n}_{i \downarrow} - \frac{1}{2} \right) 
\nonumber \\
& - \mu \sum_{i} 
\left( \tilde{n}_{i \uparrow  } - \tilde{n}_{i \downarrow} \right)
 -  \frac{h}{2}  \sum_{i} 
\left( \tilde{n}_{i \uparrow  } + \tilde{n}_{i \downarrow} \right).
\label{eq:attractive_model}
\end{align}
This reads the attractive Hubbard model 
without the spin-dependency in the transfer integrals,
therefore being free from the sign problem in the absence of 
the effective magnetic field, namely $\mu=0$~\cite{Loh_PRB1990}.
It is also understood that 
$\langle S_{\mathrm{tot}}^{z} \rangle$ is nonzero even for $h=0$,
because in terms of the attractive model,
the zero chemical potential does not correspond to
the half filling for non-bipartite lattices~\cite{DosSantos_PRB1993}.
Owing to the absence of the negative sign problem,
we can perform the QMC simulations for fairly large clusters 
with several hundreds of the lattice sites
far beyond the scope of the exact diagonalization method.
To study the bulk and boundary properties,
we treat the model under periodic boundary conditions (PBC) and 
open boundary conditions (OBC).
The total number of the unit cells $N_{\mathrm{UC}}$ is 
$L^{3}$ for PBC and $L(L+1)(L+2)/6$ for OBC,
where $L$ denotes the number of the unit cells
aligned in the linear dimension [see Fig.~\ref{fig:main}(a) for the case of OBC],
and the total number of the lattice sites $N_{\mathrm{s}}$ is $4N_{\mathrm{UC}}$.

\subsection{Phase diagram}
The model for $U>0$ has three different phases;
the HOTMI, the MI, and the correlated band insulator (cBI)
as summarized in Fig.~\ref{fig:main}(b).
Here, the cBI is the trivial band insulator with the charge and spin gaps,
thus being different from the HOTMI or the MI.
The two phase boundaries,
referred to as  $\phi_{\mathrm{c}1}^{U}$ and $\phi_{\mathrm{c}2}^{U}$,
are determined as points where 
the value of $\langle S_{\mathrm{tot}}^{z} \rangle / N_{\mathrm{UC}}$ deviates from $-1$,
which is the value of that in the HOTI or the band insulator (BI) 
at $U=0$~\cite{SM}.
This is because 
the HOTMI (cBI) is smoothly connected from the HOTI (BI)
and is therefore labeled by the same value of $\langle S_{\mathrm{tot}}^{z} \rangle$.
The phase boundaries thus determined are legitimated by calculating a more 
direct quantity, i.e., the spin gap, from magnetization plateaus under the 
nonzero magnetic field $h$~\cite{SM}.

\subsection{HOTI at $U=0$}
There are three phases at $U=0$ when $\phi$ is varied:
the HOTI, the metal, and the BI,
divided by $\phi_{\mathrm{c}1}^{0} \simeq 0.1$ and $\phi_{\mathrm{c}2}^{0} \simeq 0.4$.
In the limit of $\phi=0$ or $1/2$, 
the system is completely decoupled into a set of isolated tetrahedrons,
where the energy levels in each tetrahedron is 
$E=-3$ (3) and 1 (-1) for up (down) spin 
with the latter being threefold degenerate.
Consequently, both of the HOTI and the BI have $\langle S_{\mathrm{tot}}^{z} \rangle /N_{\mathrm{UC}}= -1$,
since the chemical potential is set as $\mu=0$.
The difference between the two gapped phases can be determined by
the $\mathbb{Z}_4$ spin-Berry phase~\cite{Hatsugai_EPL2011}: 
$\gamma=\pi$ for the HOTI and $\gamma=0$ for the BI. 
This topological invariant is defined by an integration of the many-body Berry 
connection associated with local gauge twists. 
The $S_{4}$ symmetry of the pyrochlore lattice yields its $\mathbb{Z}_4$ quantization 
as $\gamma=2\pi n/4$ with $n=0,1,2,3$~\cite{SM}.

The distinction between the HOTI and the BI can also be made by 
imposing the OBC, since according to the bulk-edge correspondence~\cite{Hatsugai_PRL1993, Hatsugai_PRB1993}
the topological property in the bulk is reflected in the edge states.
The edge states of the HOTI appear as the zero-energy states
in the energy spectra,
whereas such states are absent in the BI~\cite{Ezawa_PRL2018a,SM}.
The zero-energy states are fourfold degenerate for each spin,
originating from the isolated sites at the four corners of the finite-size 
cluster of the pyrochlore lattice [see Fig.~\ref{fig:main}(a)]
in the limit of $\phi=0$.
Therefore, the zero-energy states for $\phi < \phi_{\mathrm{c}1}^{0}$ are mostly 
localized at these corners~\cite{SM}, 
representing the third-order topological insulator 
in 3D~\cite{Ezawa_PRL2018a}.

\subsection{From HOTI to HOTMI}
\begin{figure}[tb] 
 \centering
 \includegraphics[width=1.0\linewidth]{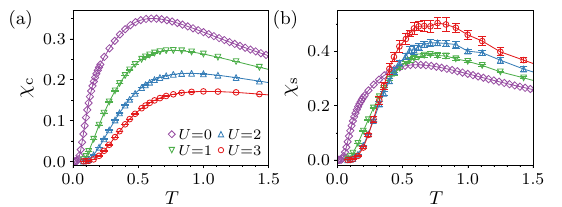}
 \caption{\label{fig:phi008-PBC}%
 Temperature dependence of 
 (a) charge compressibility $\chi_{\mathrm{c}}$ and (b) spin susceptibility $\chi_{\mathrm{s}}$
 at $\phi=0.08$ for $L=5$ under the PBC.
 }
\end{figure}

\begin{figure}[tb] 
 \centering
 \includegraphics[width=1.0\linewidth]{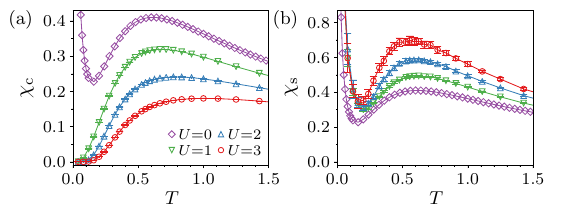}
 \includegraphics[width=1.0\linewidth]{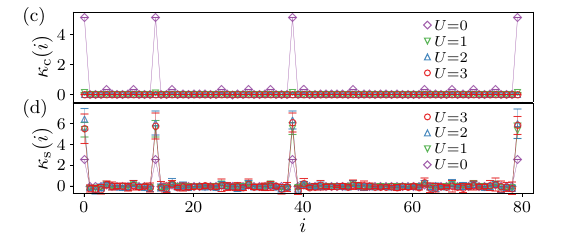}
 \caption{\label{fig:phi008-OBC}%
 Temperature dependence of 
 (a) charge compressibility $\chi_{\mathrm{c}}$ and (b) spin susceptibility $\chi_{\mathrm{s}}$
 at $\phi=0.08$ for $L=5$ under the OBC.
 Site-resolved 
 (c) charge compressibility $\kappa_{\mathrm{c}}(i)$
 and
 (d) spin susceptibility  $\kappa_{\mathrm{s}}(i)$
 for the system of $L=4$ under the OBC 
 at $\phi=0.08$ and $T=0.08$.
 }
\end{figure}

%
The HOTI changes to the HOTMI when the on-site interaction $U$ is turned on.
It is, however, difficult to distinguish these two phases 
by the bulk properties because they both have the charge and spin gaps.
In Fig.~\ref{fig:phi008-PBC}, 
we show temperature dependence of the charge compressibility $\chi_{\mathrm{c}}$
and the spin susceptibility $\chi_{\mathrm{s}}$, defined respectively as
\begin{equation}
\chi_{\mathrm{c}} = \frac{1}{N_{\mathrm{s}}} \frac{\partial \langle N \rangle}{\partial \mu} 
\end{equation}
and
\begin{equation}
\chi_{\mathrm{s}} = \frac{1}{N_{\mathrm{s}}} \frac{\partial \langle S_{\mathrm{tot}}^{z} \rangle}{\partial h},
\end{equation}
at $\phi=0.08<\phi_{\mathrm{c}1}^{0}$.
Except that $\chi_{\mathrm{c}}$ is more strongly suppressed by $U$, 
there is no obvious qualitative difference between the HOTI and the HOTMI.
On the other hand, if we consider the system under the OBC,
the difference can be noticeable 
as shown in Fig.~\ref{fig:phi008-OBC}.
At $U=0$, both $\chi_{\mathrm{c}}$ and $\chi_{\mathrm{s}}$ show a diverging behavior at low $T$,
which is due to the gapless modes in the HOTI.
For $U>0$, the gapless charge excitations vanish as shown in Fig.~\ref{fig:phi008-OBC}(a), 
whereas the gapless spin excitations remain 
as evident in the diverging behavior of $\chi_{\mathrm{s}}$ for $U>0$ [see Fig.~\ref{fig:phi008-OBC}(b)].
The feature that the boundary states posses only the charge gap 
seems common in the  
TMI~\cite{Hohenadler_review_2013, Rachel_review_2018}.

The gapless modes observed from $\chi_{\mathrm{c}}$ and $\chi_{\mathrm{s}}$ for the system under 
the OBC are elucidated by ``site-resolved'' charge compressibility 
and spin susceptibility, defined respectively as 
\begin{equation}
\kappa_{\mathrm{c}}(i) = \frac{\partial \langle n_{i} \rangle}{\partial \mu}
\end{equation}
and 
\begin{equation}
\kappa_{\mathrm{s}}(i) = \frac{\partial \langle m_{i} \rangle}{\partial h}
\end{equation}
with $m_{i} = \left( n_{i \uparrow} - n_{i \downarrow} \right)/2$,
which are similar to a momentum-resolved 
compressibility~\cite{Otsuka_PRB2002b,Otsuka_Physica2003,Morita_PRB2004}.
As shown in Fig.~\ref{fig:phi008-OBC}(c), 
$\kappa_{\mathrm{c}}(i)$ for $U=0$ exhibits peaks at four site locations
that are the isolated corners in the limit of $\phi = 0$. 
This is the expected behavior of the third-order topological insulator in three dimensions.
Note that the peaks in $\kappa_{\mathrm{s}}(i)$ of Fig.~\ref{fig:phi008-OBC}(d)
are identical to those in $\kappa_{\mathrm{c}}(i)$ (except for the constant factor) at $U=0$
because the gapless excitations appears in the single-particle spectrum.
The peaks in $\kappa_{\mathrm{c}}(i)$ immediately disappear upon inclusion of $U$, 
while the peaks in $\kappa_{\mathrm{s}}(i)$ remain and even develop for $U>0$.
This clearly shows that the gapless spin excitations appear around
the ($d-3$)-dimensional boundary, namely the corners,
which can also be observed from the enhancement of the local magnetic moments
$\langle m_{i}^{2} \rangle - \langle m_{i}^{2} \rangle_{0}$,
where $\langle \cdots \rangle_{0}$ denotes the expectation value for $U=0$,
as shown in Fig.~\ref{fig:main}(a).

It is desired
to calculate some quantity which directly characterizes 
the topological index such as the spin-Berry phase~\cite{Hatsugai_EPL2011}
for further identifying the $U>0$ phase as the HOTMI.
However, such calculation is not feasible
because there is no established way within the framework of the auxiliary-field QMC.
It is  also because the system size of the pyrochlore lattice is too large 
to apply the exact diagonalization method, 
which was possible for the kagome lattice~\cite{Kudo_PRL2019}.
Nevertheless, 
it is reasonable to consider that the nontrivial topology is protected
by the bulk charge and spin gaps as shown in Fig.~\ref{fig:phi008-PBC}.

\subsection{Collapse of the HOTMI}
\begin{figure}[tb] 
 \centering
 \includegraphics[width=1.0\linewidth]{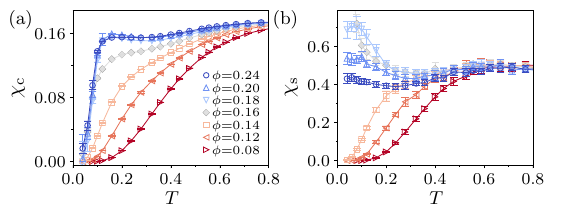}\\
 \includegraphics[width=1.0\linewidth]{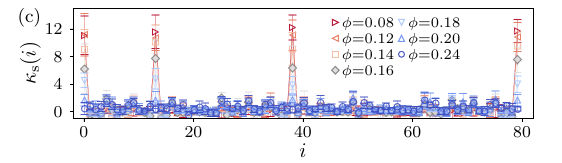}
 \caption{\label{fig:U030}%
 Temperature dependence of 
 (a) charge compressibility $\chi_{\mathrm{c}}$ and (b) spin susceptibility $\chi_{\mathrm{s}}$
 for the system of $L=5$ under the PBC  at $U=3$.
 (c) Site-resolved spin susceptibility $\kappa_{\mathrm{s}}(i)$
 for $L=4$ under the OBC at $U=3$ and $T=0.08$.
 The critical point estimated by the value of $\langle S_{\mathrm{tot}}^{z} \rangle/N_{\mathrm{UC}}$
 is $\phi_{\mathrm{c}1}^{U=3} \simeq 0.16$.
 }
\end{figure}

Next, we examine how the HOTMI evolves into the MI 
with varying $\phi$ at a fixed value of $U=3$.
We confirm in Fig.~\ref{fig:U030}(a) that the charge gap does not 
close between the HOTMI and the MI, since the temperature dependence of 
$\chi_{\mathrm{c}}$ always shows the thermally-activated behavior 
below and above $\phi_{\mathrm{c}1}^{U} \simeq 0.16$ that is determined by $\langle S_{\mathrm{tot}}^{z} \rangle/N_{\mathrm{UC}}$.
In addition, the change of $\chi_{\mathrm{c}}$ with increasing $\phi$ is found to be nonuniform.
Below $\phi_{\mathrm{c}1}^{U}$, $\chi_{\mathrm{c}}$ gradually increases as $\phi \to \phi_{\mathrm{c}1}^{U}$,
indicating that the charge gap continuously decreases.
At $\phi=\phi_{\mathrm{c}1}^{U}$, the temperature dependence of $\chi_{\mathrm{c}}$ qualitatively changes,
and they fall into the almost same curve for $\phi>\phi_{\mathrm{c}1}^{U}$, 
which suggests that the charge gap in the MI does not depend on $\phi$.
This abrupt change in $\chi_{\mathrm{c}}$ implies that the natures of the charge gaps
are different between the HOTMI and the MI.
In Fig.~\ref{fig:U030}(b), it is observed that
the thermally-activated behavior of $\chi_{\mathrm{s}}$ is completely lost for $\phi>\phi_{\mathrm{c}1}^{U}$.
The peak structure in $\kappa_{\mathrm{s}}(i)$ also vanishes when the spin gap closes at $\phi_{\mathrm{c}1}^{U}$
as shown in Fig.~\ref{fig:U030}(c)~\cite{SM}.
This topological phase transition is intrinsically different form the noninteracting 
counterpart; while in the noninteracting systems the topological property can change
when the charge and spin gaps close, here the topological phase transition occurs
when the spin gap solely closes.

\section{Discussion}

Finally,
we comment on possible realizations of the HOTMI.
The starting model of Eq.~(\ref{eq:Hamiltonian})
involves the spin-dependent transfer integrals which seems difficult to
realize in material. 
However, if we exploit the mapping of Eq.~(\ref{eq:attractive_model}),
the mapped attractive model turns out to have the hoppings which does not
dependent on the spin.
We thus expect that the HOTMI would be realized in materials 
with the breathing pyrochlore lattice structure 
and the attractive interaction at quarter filling.
Such a system may also have instability to superconductivity~\cite{large-U}.
Then, we are also tempted to speculate that some aspects of the HOTMI on the kagome 
lattice~\cite{Kudo_PRL2019} might be related to recently discovered
kagome superconductors 
AV$_{3}$Sb$_{5}$ (A=K, Rb, Cs)~\cite{Ortiz_PRM2019, Ortiz_PRM2021, Jiang_NatMat2020, Liang_PRX2021}.

We have studied the spinful Hubbard-like model on the pyrochlore lattice 
in three dimensions.
Owing to the well-designed amendment of the model, 
namely the spin-dependent transfer integrals originally proposed in the
previous study on the kagome lattice,
the model yields the higher-order topological insulator 
in the noninteracting limit.
The spin-dependent transfer integrals also enable us to study the model
by the auxiliary-field quantum Monte Carlo method, 
which is numerically exact, without suffering the negative-sign problem.
With including the interaction $U$,
we have found that 
the gapless corner spin-only excitations 
persist for the system with the open boundaries, 
while the bulk hosts both the charge and spin gaps,
which is characteristics of the topological Mott insulator.
To our best knowledge, 
this is the first unbiased evidence for 
the topological Mott insulator in three dimensions.
Furthermore, we have confirmed that 
this phase also falls within the category of 
the higher-order topological Mott insulator 
by calculating the site-resolved spin susceptibility 
showing the peaks at the corners.
The higher-order topological Mott insulator collapses into the usual 
Mott insulator when the bulk spin gap solely closes.

\section{Acknowledgments}
This work was supported by JSPS KAKENHI Grant Numbers 
JP17H06138, JP18K03475, JP18H01183, JP19J12317, JP20H04627, JP21K13850, JP21H04446, and JP21K03395.
Parts of numerical simulations have been performed on 
the HOKUSAI supercomputer at RIKEN (Project ID: G20006)
and 
the FUGAKU supercomputer provided by the RIKEN Center for Computational Science (R-CCS).

\section{Author contributions}
Y.O. developed the numerical codes, performed the QMC simulations, and 
analyzed the numerical data.
All authors conceived the project and participated in the discussion of 
the results and in the writing of the paper.

\section{Competing interests}
The authors declare no competing interests.

\section{Data availability}
The datasets generated and/or analyzed during the current study
are available from the corresponding author on reasonable request.

\bibliography{HOTMI_Pyro,HOTMI_Pyro_misc}

\clearpage

\onecolumngrid
\begin{center}
 {\Large
 \textbf{
 \textit{Supplemental Material}:\\
 Higher-Order Topological Mott Insulator on the Pyrochlore Lattice
 }
 }

\vspace{\baselineskip}

 Yuichi Otsuka,$^{1,2}$
 Tsuneya Yoshida,$^{3,4}$
 Koji Kudo,$^{4}$ 
 Seiji Yunoki,$^{1,2,5,6}$ 
 and 
 Yasuhiro Hatsugai$^{3,4}$

\vspace{\baselineskip}

 {\small
 {\it
 $^1$Computational Materials Science Research Team, 
 RIKEN Center for Computational Science (R-CCS), 
 Kobe, Hyogo 650-0047, Japan 

 $^2$Quantum Computational Science Research Team, 
 RIKEN Center for Quantum Computing (RQC), 
 Wako, Saitama 351-0198, Japan

 $^3$Graduate School of Pure and Applied Sciences, 
 University of Tsukuba, Tsukuba, Ibaraki 305-8571, Japan

 $^4$Department of Physics,
 University of Tsukuba, Tsukuba, Ibaraki 305-8571, Japan

 $^5$Computational Condensed Matter Physics Laboratory, 
 RIKEN, Wako, Saitama 351-0198, Japan

 $^6$Computational Quantum Matter Research Team, 
 RIKEN Center for Emergent Matter Science (CEMS), 
 Wako, Saitama 351-0198, Japan

 }
 }
\end{center}
\vspace{\baselineskip}

\twocolumngrid

\renewcommand{\thesection}{S\arabic{section}}
\renewcommand{\theequation}{S\arabic{equation}}
\renewcommand{\thefigure}{S\arabic{figure}}
\renewcommand{\thetable}{S\arabic{table}}
\setcounter{equation}{0}
\setcounter{figure}{0}
\setcounter{table}{0}
\setcounter{page}{1}
\setcounter{enumi}{0} 
\renewcommand{\bibnumfmt}[1]{[S#1]}
\renewcommand{\citenumfont}[1]{S#1}
\makeatletter
\c@secnumdepth = 2

\section{$\mathbb{Z}_4$ spin-Berry phase}
\label{Sec:Z4BP}

\begin{figure}[tb]
 \centering
 \includegraphics[width=1.0\linewidth]{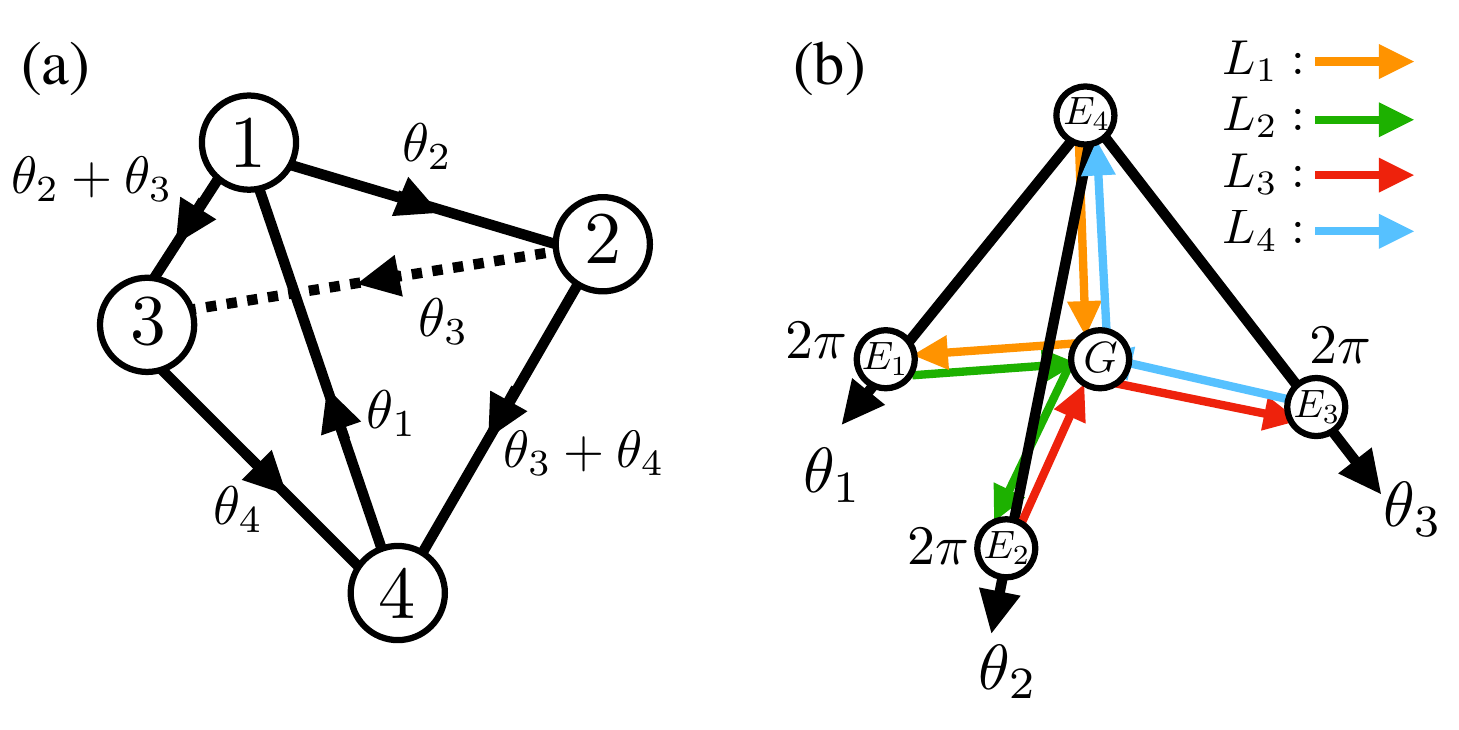}
 \caption{\label{fig:Py}%
 (a) Peierls phase in a downward tetrahedron of the pyrochlore lattice. 
 (b) Integration paths $L_1,\cdots,L_4$. $G$ represents the center of 
 gravity.}
 \end{figure}

In this appendix, we describe the definition of the spin-Berry phase $\gamma$
and discuss its $\mathbb{Z}_4$ quantization~\cite{Hatsugai_EPL2011SM,KK-thesisSM}.
The spin-Berry phase is given by an integration over the angles of local gauge
twist defined below. Let us consider the bulk system on the pyrochlore lattice.
Picking up a
specific downward tetrahedron, we define a unitary operator as
\begin{align}
  U_-(\vec{\theta})=\exp\left\{i\sum_{j=1}^4n_j^-\phi_j\right\},
\end{align}
where $j=1,\cdots,4$ is the site index of the tetrahedron, 
$n_j^-=n_{j\uparrow}-n_{j\downarrow}$, $\phi_j=\sum_{k=1}^j\theta_k$, and
$\vec{\theta}=(\theta_1,\theta_2,\theta_3,\theta_4)$ is a four-dimensional
parameter defined 
on a torus $T^4$. Setting $\mu=h=0$ in 
Eq.~(1), we then rewrite the Hamiltonian as 
\begin{align}
 \mathcal{H}(\vec{\theta})
 =\mathcal{H}^\bigtriangleup_t
 +U_-(\vec{\theta})\mathcal{H}_t^{\bigtriangledown}U_-^\dagger(\vec{\theta})
 +\mathcal{H}_U.
\end{align}
This modification brings the Peierls phase in the hopping term within the 
chosen downward tetrahedron as shown in Fig.~\ref{fig:Py}(a). Let us 
define the spin-Berry connection as
\begin{align}
 \vec{A}(\vec{\theta})
 =\bra{G(\vec{\theta})}\vec{\nabla}_{\vec{\theta}}
 \ket{G(\vec{\theta})},
\end{align}
where 
$\vec{\nabla}_{\vec{\theta}}=(\partial/\partial\theta_1,\cdots,\partial/\partial\theta_4)$ 
and $|G(\vec{\theta})\rangle$ is the ground state of 
$\mathcal{H}(\vec{\theta})$. The spin-Berry phase is defined as
\begin{align}
 \gamma_i
 =\frac{1}{i}\int_{L_i}d\vec{\theta}\cdot\vec{A}(\vec{\theta}).
 \label{eq:bp}
\end{align}
The integration path $L_i$ is given as follows. Defining five
points in the parameter space as 
\begin{align*}
 &E_1=(2\pi,0,0,0),\\ 
 &E_2=(0,2\pi,0,0),\\
 &E_3=(0,0,2\pi,0),\\ 
 &E_4=(0,0,0,2\pi),\\
 &G=\frac{1}{4}(2\pi,2\pi,2\pi,2\pi),
\end{align*}
we introduce paths from $E_i$ to $G$ as shown in Fig.~\ref{fig:Py}(b) by 
setting $\phi_4=2\pi$, i.e., $\theta_4=2\pi-\theta_1-\theta_2-\theta_3$. They are 
expressed as~\cite{ArakiSM}
\begin{align*}
 &\vec{f}_1(t)=\frac{2\pi}{4}(4-3t,t,t,t),\\ 
 &\vec{f}_2(t)=\frac{2\pi}{4}(t,4-3t,t,t),\\ 
 &\vec{f}_3(t)=\frac{2\pi}{4}(t,t,4-3t,t),\\ 
 &\vec{f}_4(t)=\frac{2\pi}{4}(t,t,t,4-3t),   
\end{align*}
where $0\leq t\leq1$. Along these lines, the integral path is
defined as $L_i:E_{i-1}\rightarrow G\rightarrow E_{i}$, where $E_0\equiv E_4$.

\begin{figure}[t]
   \begin{center}
    \includegraphics[width=\columnwidth]{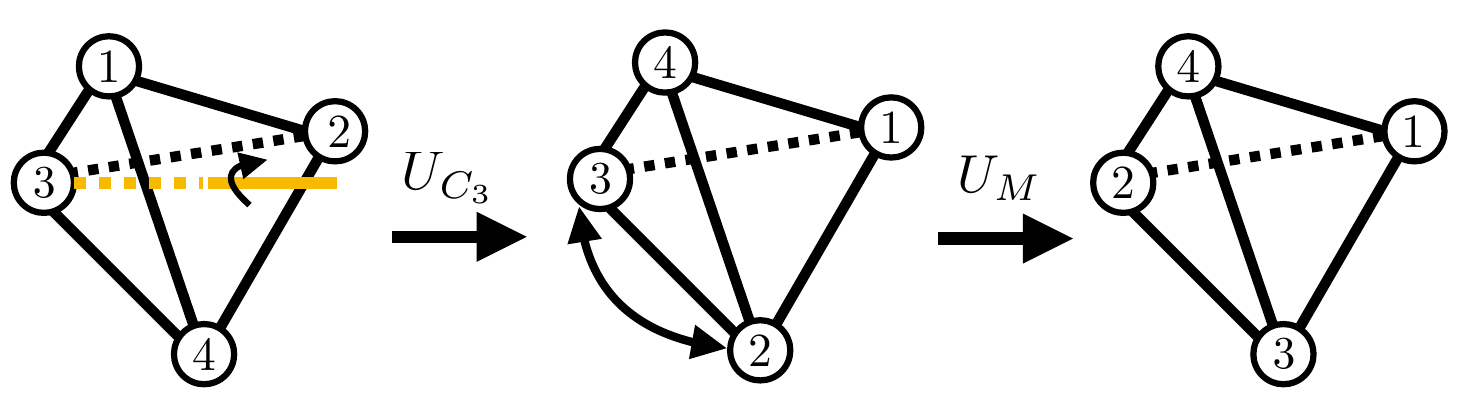}
   \end{center}
 \caption{\label{fig:sym}%
Unitary operators $U_{C_3}$ and $U_M$.}
\end{figure}

Due to the equivalence of the paths $L_1,\cdots,L_4$, the spin-Berry phase is 
quantized into $\mathbb{Z}_4$ . Let us now derive it in detail. 
Tetrahedron is invariant for any exchange of the vertexes by the symmetric 
group $S_4$~\cite{Hatsugai_EPL2011SM}. This implies that the original Hamiltonian 
$\mathcal{H}(\vec{\theta}=\vec{0})$ have $C_3$ symmetry and
mirror symmetry with respect to the chosen downward tetrahedron, see
Fig.~\ref{fig:sym}. With finite 
$\vec{\theta}$, the symmetry is broken but instead we have 
\begin{align}
 (U_MU_{C_3})H(\vec{\theta})(U_MU_{C_3})^{-1}
 =H(g\vec{\theta}),
 \label{eq:sym}
\end{align}
where $g$ is the $4\times4$ unitary matrix satisfying 
$g\vec{\theta}=(\theta_4,\theta_1,\theta_2,\theta_3)$. 
Clearly, we have
\begin{align}
 g\vec{f}_{i}(t)=\vec{f}_{i+1}(t),
 \label{eq:cyc}
\end{align} 
where $i=1,\cdots,4$ and $\vec{f}_5\equiv\vec{f}_1$. 
This implies 
$\gamma_1\equiv\gamma_2\equiv\gamma_3\equiv\gamma_4\equiv\gamma\mod2\pi$ as 
follows:
\begin{align*}
 \gamma_{i}
 &=\sum_j\frac{1}{i}\int_{L_{i}}d\theta_j\bra{ 
 G(\vec{\theta})}\frac{\partial}{\partial\theta_j}\ket{G(\vec{\theta})}\\
 &=\sum_{j}\frac{1}{i}\int_{L_{i-1}}d\left(\sum_kg_{jk}\theta_k'\right)\times\\
 &\qquad\qquad
 \bra{G(g\vec{\theta}')}\sum_l(g^{-1})_{lj}
 \frac{\partial}{\partial\theta_l'}\ket{G(g\vec{\theta}')}\\
 &=\sum_{jkl}g_{jk}(g^{-1})_{lj}\frac{1}{i}\int_{L_{i-1}}d\theta_k'\bra{ 
 G(g\vec{\theta}')}
 \frac{\partial}{\partial\theta_l'}\ket{G(g\vec{\theta}')}\\
 &=\sum_{l}\frac{1}{i}\int_{L_{i-1}}d\theta_l'\bra{ 
 G(g\vec{\theta}')}\frac{\partial}{\partial\theta_l'}\ket{G(g\vec{\theta}')}\\
 &=\sum_{l}\frac{1}{i}\int_{L_{i-1}}d\theta_l'\bra{ 
 G(\vec{\theta}')}\frac{\partial}{\partial\theta_l'}\ket{G(\vec{\theta}')}\\
 &=\gamma_{i-1},
\end{align*}
where $\vec{\theta}=g\vec{\theta}'$,
and we use
$\partial/(\partial\theta_l')\ket{G(g\vec{\theta}')}
=\left(U_MU_{C_3}\right)^{-1}\partial/(\partial\theta_l')
\ket{G(\vec{\theta}')}$.
Since the sum of the loop $L_1,\cdots,L_4$ is equal to zero, 
implying
\begin{align}
 \sum_i\gamma_i\equiv0 \mod2\pi,
\end{align}
we have 
\begin{align}
 \gamma\equiv\frac{n}{4}2\pi
 \mod2\pi,
 \label{eq:z4}
\end{align}
where $n=0,1,2,3$. 

Since the quantized value does not change unless the energy gap closes, 
$\gamma$ is an adiabatic invariant for gapped topological phases. For $U=0$, 
we have $\gamma=\pi$ for the HOTI while $\gamma=0$ for the band insulator. Let 
us now demonstrate it based on the decoupled limit. The HOTI phase includes
the decoupled system with $t_\bigtriangleup=0$, whose Hamiltonian is given by
$\mathcal{H}(\vec{\theta})
=U_-(\vec{\theta})\mathcal{H}_t^{\bigtriangledown}
U_-^\dagger(\vec{\theta})$.
The spin-Berry connection $\vec{A}=(A_1,\cdots,A_4)$ is calculated as 
\begin{align*}
 A_j(\vec{\theta})=
 \bra{G(\vec{\theta})}\frac{\partial}{\partial\theta_j}
 \ket{G(\vec{\theta})}
 &=\bra{G_0}U_-^\dagger(\vec{\theta})\frac{\partial}{\partial\theta_j}
 U_-(\vec{\theta})\ket{G_0}\\
 &=\bra{G_0}\left(i\sum_{k=j}^4n_k^-\right)\ket{G_0},
\end{align*}
where $\ket{G_0}=\ket{G(\vec{\theta}=\vec{0})}$. 
Because of symmetry, we have 
$\bra{G_0}n_1^-\ket{G_0}=\cdots=\bra{G_0}n_4^-\ket{G_0}\equiv s$ and
\begin{align}
 A_j(\vec{\theta})
 &=i(5-j)s.
 \label{eq:A}
\end{align}
Consequently, the $\mathbb{Z}_4$ spin-Berry phase is given by
\begin{align}
 \gamma_1
 &=\frac{1}{i}\int_{L_1}d\vec{\theta}\cdot \vec{A}(\vec{\theta})\nonumber\\
 &=\frac{1}{i}\int_0^{2\pi}d\theta_1A_1(\vec{\theta})
 +\frac{1}{i}\int_{2\pi}^{0}d\theta_4A_4(\vec{\theta})\nonumber\\
 &=\frac{1}{i}\int_0^{2\pi}d\theta_1\left(i4s\right)
 +\frac{1}{i}\int_{2\pi}^{0}d\theta_4\left(is\right)\nonumber\\
 &=6\pi s.
\end{align}
As mentioned in the main text, we have $s=-1/2$ in the half-filling, which 
implies $\gamma=\pi\mod2\pi$.
The other limit, i.e., $t_\bigtriangledown=0$ is included in the band 
insulating phase. Its Hamiltonian is given by 
$H(\vec{\theta})=H^\bigtriangleup$. Because of the independence of 
$\vec{\theta}$, we have $\gamma=0 \mod2\pi$.

\section{Computational details of QMC simulations}
\label{Sec:computatinal}

In the scheme of the auxiliary field QMC,
the Suzuki-Trotter decomposition~\cite{Suzuki_1976SM,Trotter_1959SM} is
first applied to  $e^{-\beta \mathcal{H}}$ as 
$e^{-\beta \mathcal{H}} \simeq \prod 
e^{-\Delta \tau \mathcal{H}_{t}/2} 
e^{-\Delta \tau \mathcal{H}_{U}}  
e^{-\Delta \tau \mathcal{H}_{t}/2}$,
where $\mathcal{H}_{t}$ stands for the noninteracting parts 
in $\mathcal{H}$, and
$\Delta \tau = \beta/M$ is a Trotter slice with $M$ being integer.
The discrete Hubbard-Stratonovich transformation~\cite{Hirsch_PRB1983SM} is then
applied to each term of $e^{-\Delta \tau U n_{i \uparrow} n_{i \downarrow}}$,
introducing an auxiliary Ising-type variable 
at each spatial site in each imaginary time slice.
The summation over the auxiliary fields involving the $M N_{\mathrm{s}}$ Ising
variables is performed by Monte Carlo (MC) sampling.

We set the Trotter slice as $\Delta \tau =0.1$, 
for which the Trotter errors of order $O(\Delta \tau^{2})$
are sufficiently small compared with statistical errors of the MC sampling.
As for the Hubbard-Stratonovich transformation,
we employ one which couples to the spin degree of freedom.
Typically, we perform $4 \times 10^{3}$ MC sweeps for equilibration,
followed by $8 \times 10^{4}$ MC sweeps for measurement, 
which are divided into 20 bins to estimate the statistical error
by the standard deviation.
Each MC sweep consists of $M N_{\mathrm{s}}$ local updates and
$N_{\mathrm{s}}$ global moves~\cite{Scalettar_PRB1991SM}.
The simulations are carried out on the finite size clusters
with $L$ up to 5 (8) corresponding to $N_{\mathrm{s}}=500$ (480)
under the PBC (OBC).

\section{Energy spectra at $U=0$}
\label{Sec:spectra}

\begin{figure}[tb] 
 \centering
 \includegraphics[width=0.49\linewidth]{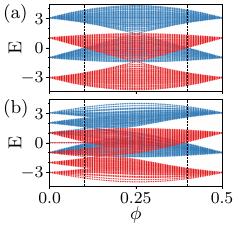}
 \includegraphics[width=0.49\linewidth]{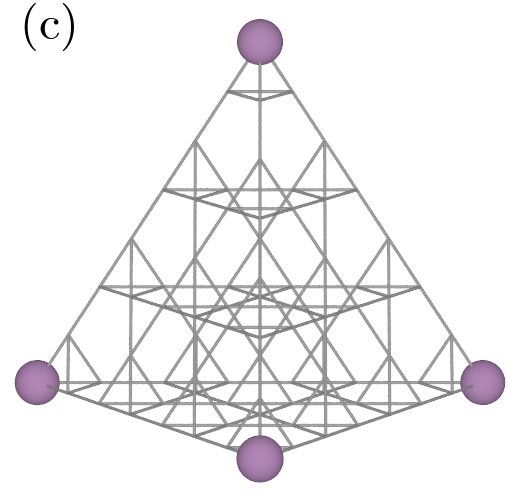}
 \caption{\label{fig:spectra}%
 Energy spectra as function of $\phi$ at $U=0$ for 
 (a) $L=10$ under the PBC and
 (b) $L= 8$ under the OBC. 
 Red (blue) symbols represent the energy levels for up (down) spin. 
 Vertical dashed lines indicate
 $\phic{0}{1} \simeq 0.1$ and $\phic{0}{2} \simeq 0.4$.
 (c) The averaged probability densities of the degenerated zero-energy 
 states for $L=4$ under the OBC are shown by the radius of the purple
 spheres for $\phi=0.04$.
 }
\end{figure}

The energy spectra as function of $\phi$ in the noninteracting limit 
are shown in Fig.~\ref{fig:spectra}.
For the system under the PBC, the single-particle gap opens for
the HOTI of $\phi<\phic{0}{1} \simeq 0.1$ and 
the BI   of $\phi>\phic{0}{2} \simeq 0.4$.
For the system under the OBC, the eightfold degenerate zero-energy states
appear only in the HOTI. 
The averaged probability densities of these degenerated states are shown 
in  Fig.~\ref{fig:spectra}(c) for $\phi=0.04$.

\section{Determination of phase boundaries}
\label{Sec:Sz-T}

\begin{figure}[tb] 
 \centering
 \includegraphics[width=1.0\linewidth]{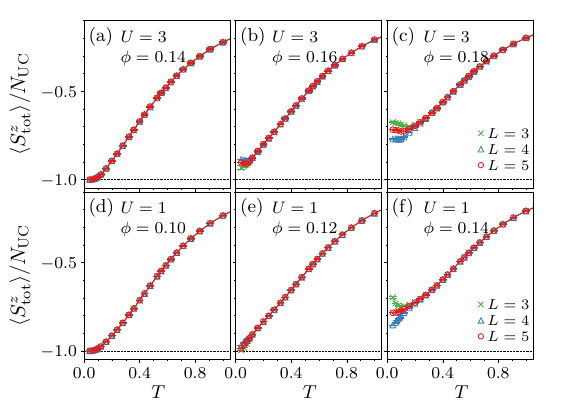}
 \caption{\label{fig:Sz-T}%
 Temperature dependence of $\szt / \nuc$ 
 for various values of $\phi$ under the PBC.
 Upper [(a)-(c)] and lower [(d)-(f)] panels show
 the results of $U=3$ and $U=1$, respectively.
 The horizontal dashed lines indicate $\szt / \nuc=-1$,
 which is the value of the ground-state in the HOTMI phase.
 }
\end{figure}

As shown in Fig.~\ref{fig:Sz-T}, 
we find $\phic{U}{1}=0.15(1)$ and $0.13(1)$ for $U=3$ and $1$.
Above $\phic{U}{1}$, the strong finite-size effect is observed at low $T$
[see Figs.~\ref{fig:Sz-T}(c) and \ref{fig:Sz-T}(f)],
implying the absence of the bulk spin gap.

\section{Collapse of spin gap}
\label{Sec:spin_gap}

\begin{figure}[tb] 
 \centering
 \includegraphics[width=1.00\linewidth]{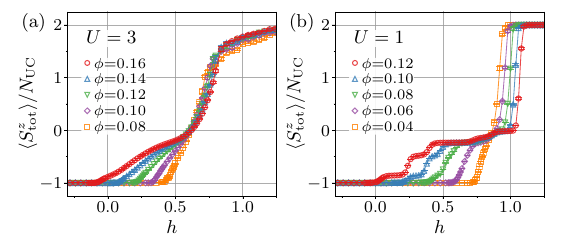}\\
 \includegraphics[width=1.00\linewidth]{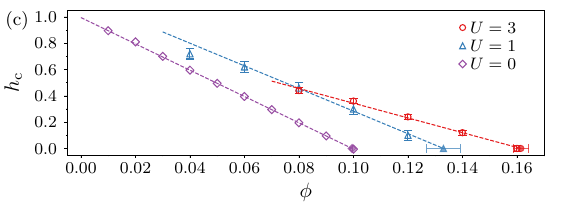}
 \caption{\label{fig:Sz-h}%
 $\szt / \nuc$ as a function of the magnetic field $h$
 for (a) $U=3$ and (b) $U=1$,
 and $L=4$ under the PBC at $T=0.025$,
 from which the critical magnetic fields $h_{\mathrm{c}}$ are determined.
 (c) $\phi$-dependence of $h_{\mathrm{c}}$ for $U=3$ and $1$.
 For comparison, corresponding exact values for $U=0$ are also shown.
 The dashed lines are linear fits to the data points.
 The filled symbols at $h_{\mathrm{c}}=0$ indicate $\phic{U}{1}$
 estimated from the fittings.
 }
\end{figure}

We confirm that the spin gap indeed vanishes at $\phi=\phic{U}{1}$ 
from magnetization plateaus under the magnetic field $h$ as shown 
in Fig.~\ref{fig:Sz-h}.
It is noted that the simulations for $h \neq 0$
are possible without encountering the negative sign problem,
since the model can be mapped onto the attractive model.
The critical magnetic field, $h_{\mathrm{c}}$, is determined,
in the similar way to $\phic{U}{1}$,
as the point above which $\szt / \nuc$ deviates from $-1$.
Since the spin gap is proportional to $h_{\mathrm{c}}$, 
the $\phi$-dependence of $h_{\mathrm{c}}$ in Fig.~\ref{fig:Sz-h}(c)
represents how the spin gap decreases.
Thus, the critical point $\phic{U}{1}$ is estimated as the point of $\phi$
for which $h_{\mathrm{c}}$ is zero.
The result in Fig.~\ref{fig:Sz-h}(c) shows that
$\phic{U}{1}$ estimated in this way turn out to agree well with 
those obtained from the $T$-dependence of $\szt / \nuc$
in Fig.~\ref{fig:Sz-T} within the error bars.

\end{document}